\documentclass[cite]{cimento}

\usepackage{epsfig}
\usepackage{amsmath}
\usepackage{booktabs}
\usepackage{xspace}

\newcommand{\bbar}{\ensuremath{\bar{b}}\xspace}
\newcommand{\tbar}{\ensuremath{\bar{t}}\xspace}
\newcommand{\qbar}{\ensuremath{\bar{q}}\xspace}
\newcommand{\bbbar}{\ensuremath{{b\bbar}}\xspace}
\newcommand{\ttbar}{\ensuremath{{t\tbar}}\xspace}
\newcommand{\qqbar}{\ensuremath{{q\qbar}}\xspace}
\newcommand{\tZq}{\ensuremath{{t\rightarrow Zq}}\xspace}
\newcommand{\costh}{\ensuremath{\cos\theta^*}\xspace}
\newcommand{\pt}{\ensuremath{p_{T}}\xspace}

\newcommand{\ptb}{\ensuremath{p_{T}(b)}\xspace}
\newcommand{\pyth}{{\sc Pythia}\xspace}
\newcommand{\herwig}{HERWIG\xspace}
\newcommand{\madevent}{\texttt{MadEvent}\xspace}
\newcommand{\VminusA}{\ensuremath{V\!-\!A}\xspace}
\newcommand{\alp}{ALPGEN\xspace}
\newcommand{\dzero}{D\O\xspace}
\newcommand{\ie}{\emph{i.e.}\ }
\newcommand{\eg}{\emph{e.g.}\ }
\newcommand{\unit}[1]{\ensuremath{\:\mathrm{#1}}\xspace}
\newcommand{\invfb}{\unit{fb^{-1}}}

\newcommand{\gevc}{\ensuremath{\unit{GeV}/c}\xspace}

\newcommand{\tev}{\unit{TeV}}

\title{Monte Carlo Simulations for Top Pair and Single Top Production
  at the Tevatron} 

\shorttitle{MC Simulations for Top Pair and Single Top Production at
  the Tevatron}

\author{U.~Husemann\from{zeuthen} on behalf of the CDF and \dzero
  Collaborations}\shortauthor{U.~Husemann}

\instlist{\inst{zeuthen} Deutsches Elektronen-Synchrotron,
  Platanenallee~6, 15738~Zeuthen, Germany}

\PACSes{
  \PACSit{14.65.Ha}{Top quarks}
  \PACSit{2.70.Uu}{Applications of Monte Carlo methods}
}

\begin{document}

\maketitle

%\vspace{-80mm}
%
%\begin{flushright}
%CDF/PUB/TOP/PUBLIC/9369\\
%FERMILAB-CONF-08-191-E
%end{flushright}
%\vspace{70mm}

%%%%%%%%%%%%%%%%%%%%%%%%%%%%%%%%%%%%%%%%%%%%%%%%%%%%%%%%%%%%%%%%%%%%%%%%
%
% Abstract
%
%%%%%%%%%%%%%%%%%%%%%%%%%%%%%%%%%%%%%%%%%%%%%%%%%%%%%%%%%%%%%%%%%%%%%%%%

\begin{abstract}
  Monte Carlo (MC) simulations are indispensable tools for top quark
  physics, both at the current Tevatron collider and the upcoming
  Large Hadron Collider. In this paper we review how the Tevatron
  experiments CDF and \dzero utilize MC simulations for top quark
  analyses. We describe the standard MC generators used to simulate
  top quark pair and single top quark production, followed by a
  discussion of methods to extract systematic uncertainties of top
  physics results related to the MC generator choice. The paper also
  shows the special MC requirements for some example top properties
  measurements at the Tevatron.
\end{abstract}

%%%%%%%%%%%%%%%%%%%%%%%%%%%%%%%%%%%%%%%%%%%%%%%%%%%%%%%%%%%%%%%%%%%%%%%%
%
% Introduction
%
%%%%%%%%%%%%%%%%%%%%%%%%%%%%%%%%%%%%%%%%%%%%%%%%%%%%%%%%%%%%%%%%%%%%%%%%

\section{Introduction}

With more than 3.5\invfb of data recorded per Tevatron experiment to
date (Summer~2008), top quark physics has entered an era of precision
measurements. The latest combination of top quark mass measurements
from the CDF and \dzero collaborations shows a relative uncertainty of
only 0.8\%~\cite{TopMass2008}, and multivariate analysis techniques
are being used in many analyses, notably the recent evidence for the
electroweak production of single top
quarks~\cite{SingleTopD0,SingleTopCDF}. For all these endeavors, Monte
Carlo (MC) simulations are key tools, for estimating the signal
acceptance and the amount of background, and for evaluating systematic
uncertainties.

%%%%%%%%%%%%%%%%%%%%%%%%%%%%%%%%%%%%%%%%%%%%%%%%%%%%%%%%%%%%%%%%%%%%%%%%
%
% MC Generators
%
%%%%%%%%%%%%%%%%%%%%%%%%%%%%%%%%%%%%%%%%%%%%%%%%%%%%%%%%%%%%%%%%%%%%%%%%

\section{Standard Monte Carlo Generators at the Tevatron}
\label{section:mcgen}

\subsection{Top Pair Production}

The Tevatron collaborations follow different approaches for their
default MC generators for top quark pair production, as summarized in
table~\ref{table:defaultMC}. Both collaborations use leading order
(LO) MC generators with parton showering (PS), \pyth
v6.2~\cite{Pythia} in the case of CDF, and \alp v2.1~\cite{ALPGEN}
(using \pyth~v6.3 for PS) for \dzero. As both MC generators are LO
generators, the generated \ttbar production cross sections must be
scaled to the theoretical cross section. CDF has opted for a well
established MC generator that has been carefully validated against and
tuned to the Tevatron data. \dzero uses a fairly recent MC generator
with additional functionality compared to \pyth, including exact
matrix elements for $2\to n$ processes with a matching procedure
between partons generated via matrix elements and parton showers, a
more recent set of parton distribution functions (PDFs), and \ttbar
spin correlations. The collaborations have also chosen different ways
of simulating multiple collisions in the same bunch crossing.  CDF
overlays minimum bias events generated with \pyth, with the same
calibration constants applied as for the data events. \dzero obtains a
good description of beam and instrumental backgrounds when overlaying
events recorded with a zero bias (\ie random) trigger in data.

\begin{table}[t]
  \caption{Comparison of the default MC generators used for top quark
    pair production by the CDF and \dzero collaborations.}
\label{table:defaultMC}

\begin{tabular}{lll}
  \toprule
  & {\bf CDF} & {\bf \dzero} \\
  \midrule
  Generator & \pyth v6.2 & \alp v2.1 with \pyth v6.3\\
  Process   & $\qqbar,gg\to\ttbar$ & $\qqbar,gg\to\ttbar$\,+\,0--2 partons\\
  Parton Distribution Functions & CTEQ5L~\cite{CTEQ5} & CTEQ6L~\cite{CTEQ6}\\
  Tunes     & Tune A, $W/Z\,\,\pt$ & --- \\
  Multiple Collisions & Minimum Bias (\pyth) & Zero Bias (Data)\\
  \bottomrule
\end{tabular}

\end{table}

\subsection{Single Top Production}

The Tevatron collaborations have reported evidence for electroweak
single top quark production~\cite{SingleTopD0,SingleTopCDF}. At
$\sqrt{s} = 1.96\tev$ single top quarks are predominantly produced via
the $t$-channel process $qb \to q't$ and the $s$-channel process $q
\qbar' \to t\bbar$. Theoretical predictions of the production cross
section are available at next-to-leading order (NLO), see \eg
ref.~\cite{ZTOP}.

The LO kinematics of the $s$-channel process are unchanged by NLO
corrections; therefore it is sufficient to employ a LO MC~generator to
simulate the $s$-channel process and to scale the obtained cross
section to the NLO expectation. On the other hand there are important
corrections to the $t$-channel kinematics from the $2\to3$ process
$qg\to q't\bbar$, where the gluon in the initial state produces a
\bbbar pair. The soft and collinear regime of this process is well
modeled by the $2\to2$ process available in LO MC with PS, using $b$
quark PDFs; however, for large transverse momenta of the final state
\bbar quark, \ptb, the $2\to3$ process needs to be considered
explicitly.

\begin{figure}[t]
  \begin{center}
    \includegraphics[width=0.45\textwidth]{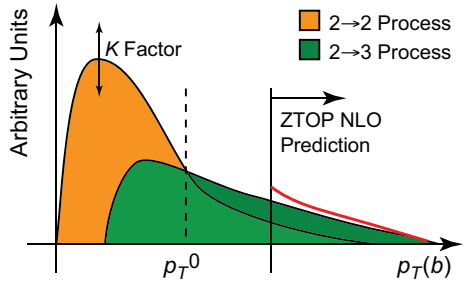}
\end{center}
\caption{Constructing the top $t$-channel cross section as a function
  of the \bbar quark transverse momentum from the $2\to2$ and the
  $2\to3$ process~\cite{Boos,Lueck,TeV4LHC}. }
\label{fig:singletop}
\end{figure}

\begin{figure}[t]
  \begin{center}
    \includegraphics[height=40mm]{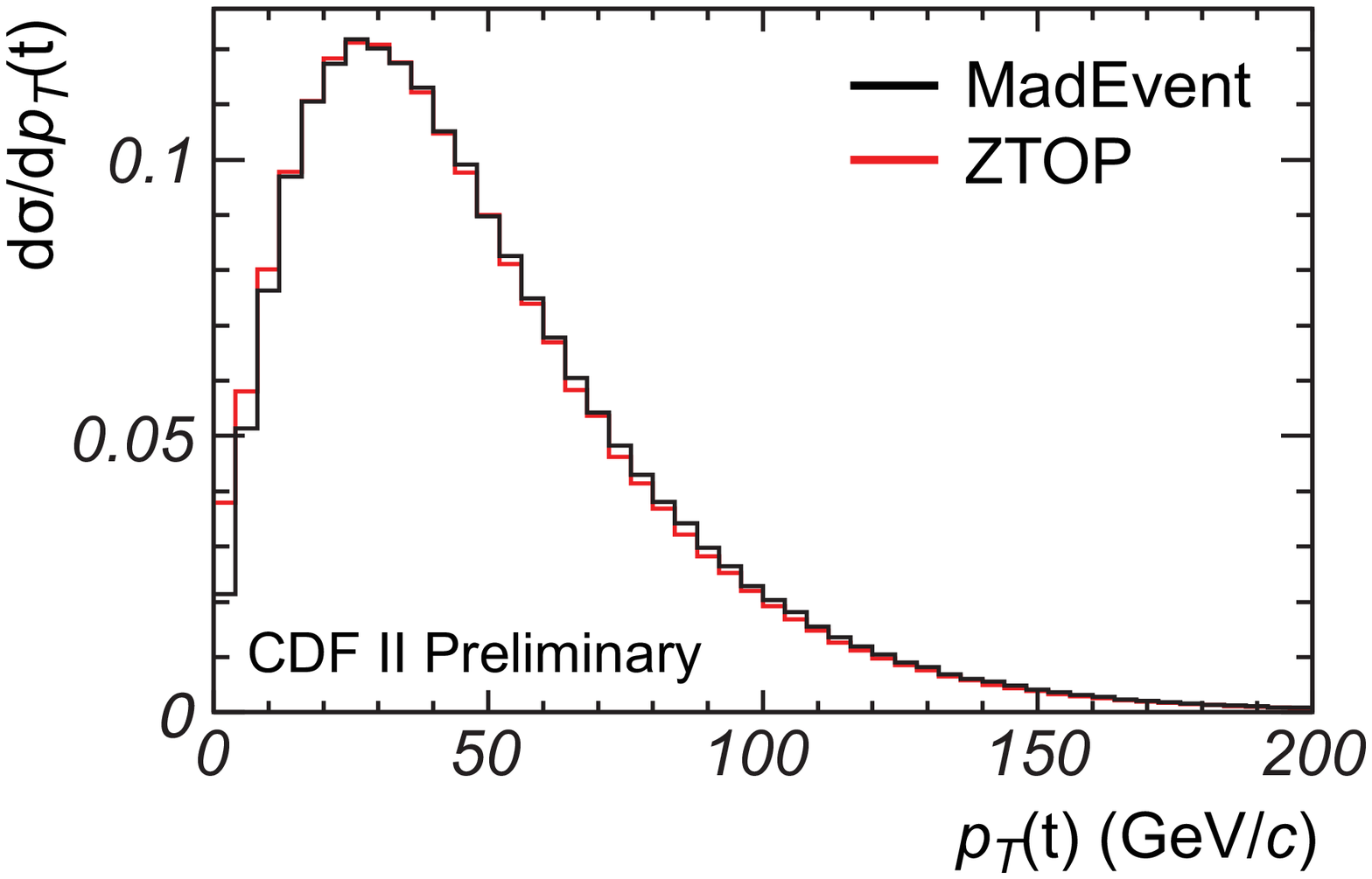}
    \hspace{5mm}
    \includegraphics[height=43mm]{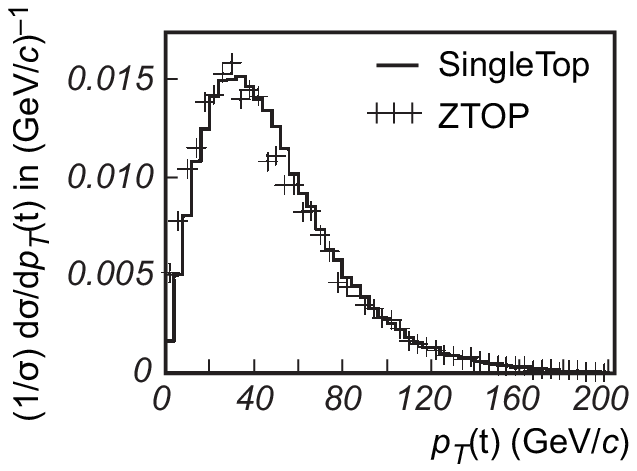}
  \end{center}
  \caption{Comparison of top quark transverse momentum distributions
    in $t$-channel single top production. Left: CDF comparison of
    \texttt{MadGraph} and ZTOP.  Right: \dzero comparison of SingleTop
    and ZTOP (after ref.~\cite{Boos}).}
\label{fig:pt_top}
\end{figure}

Both CDF and \dzero use a procedure in which the $2\to2$ and the
$2\to3$ processes are generated separately and the phase space overlap
between the two processes is removed by hand~\cite{Boos}. The NLO
cross section $\sigma_\mathrm{NLO}$ as a function of \ptb is then
constructed as follows:
\begin{equation}
  \sigma_\mathrm{NLO} = \left. K
    \cdot \sigma_{2\to2, \text{\sc Pythia}} \right|_{p_T(b)<p_T^0} +
  \left. \sigma_{2\to 3} \right|_{p_T(b)\ge p_T^0},
\end{equation}
which is illustrated in fig.~\ref{fig:singletop} and implies the
following steps:
\begin{itemize}
\item The $2\to2$ and the $2\to3$ processes are generated separately,
  utilizing the \madevent generator~\cite{MadEvent} in the case of
  CDF, and the SingleTop generator~\cite{Boos} at \dzero.
\item The $2\to2$ process is scaled up by a factor of $K$ such that
  the cross section for both processes matches the NLO cross section.
\item The soft and the hard regime are separated by a simple cut on
  \ptb at $\pt^0$, \ie all events with $\ptb<\pt^0$ are taken from the
  $2\to2$ process and all events with $\ptb>\pt^0$ are taken from the
  $2\to3$ process. The value of $\pt^0$ is chosen to ensure a smooth
  transition of the differential cross section as a function of \ptb
  between the soft and the hard regime (around
  $20\gevc$)~\cite{Lueck,TeV4LHC}.
\item The full event kinematics are then validated against NLO
  predictions from the ZTOP code~\cite{ZTOP}.  As shown in
  fig.~\ref{fig:pt_top}, both collaborations find good agreement
  between the MC samples and the NLO prediction.
\end{itemize}

The above procedure represent a ``pragmatic'' solution that is
sufficient for the current precision of the single top measurements.
In the future it is desirable to treat $t$-channel single top production
consistently in a full NLO MC simulation, \eg in the framework
of MC@NLO~\cite{MCATNLO}.

%%%%%%%%%%%%%%%%%%%%%%%%%%%%%%%%%%%%%%%%%%%%%%%%%%%%%%%%%%%%%%%%%%%%%%%%
%
% Systematic uncertainties
%
%%%%%%%%%%%%%%%%%%%%%%%%%%%%%%%%%%%%%%%%%%%%%%%%%%%%%%%%%%%%%%%%%%%%%%%%

\section{Monte Carlo Simulations and Systematic Uncertainties}

CDF and \dzero have established their (separate) ways of assigning
systematic uncertainties to top physics results over the course of
Tevatron Runs~I and~II. A significant fraction of the systematic
uncertainties is related to the MC simulation. For example, for the
most recent Tevatron top mass combination, out of the total relative
uncertainty of 0.8\%, 0.3\% can be attributed to MC related
effects~\cite{TopMass2008}. In the following we will review the
standard treatment of MC related systematic uncertainties by
CDF and \dzero and comment on recent progress on this subject.

\subsection{Signal Model}

The two main all-purpose LO MC generators with PS, \pyth~\cite{Pythia}
and \herwig~\cite{Herwig}, have different hadronization models and
different tunings for the underlying event (\ie interactions of
partons inside the colliding proton and antiproton other than the ones
involved in the hard scattering process).  The treatment of initial
state radiation (ISR) and final state radiation (FSR) in \pyth and
\herwig is similar, but not identical.  

CDF extracts MC model uncertainties for many observables from a
comparison of \pyth and \herwig. For the first years of Tevatron
Run~II, the \dzero collaboration did not employ the \herwig MC
generator and therefore did not compare \herwig to \pyth.  Instead,
\dzero evaluates uncertainties due to the $b$~quark fragmentation
model, a major source of MC model uncertainty in $b$~rich top events.

CDF and \dzero also follow different approaches to evaluating
uncertainties due to ISR and FSR.  CDF estimates the uncertainty from
special MC samples in which the \pyth parameters controlling the
amount of ISR and FSR ($\Lambda_\mathrm{QCD}$ and $Q^2$ scales) are
varied. This procedure is sensitive to differences in the soft
radiation. \dzero extracts the systematic uncertainties due to ISR and
FSR from reweighting the jet multiplicity spectrum in the MC
simulation (\alp $\ttbar +$\,0--2\,partons) in a control region to the
multiplicity spectrum observed in data. With this procedure, \dzero
mainly probes differences in hard radiation.

Both collaboration evaluate uncertainties due to the PDF choice in
similar ways. For any given observable, \eg the acceptance for \ttbar
events, the uncertainty is obtained from two main sources, the change
in the observable when varying eigenvectors in the space of the PDF
fit parameters and the difference in the observable when calculated
using CTEQ~\cite{CTEQ6} versus MRST~\cite{MRST} PDF sets.

\subsection{Jet Energy Scale}

In many top analyses, uncertainties in the calibration of the jet
energy scale (JES) are among the leading systematic uncertainties. The
CDF procedure for JES corrections is detailed in ref.~\cite{JESCDF}.
The JES correction is obtained \eg from dijet and photon-jet balance,
and the uncertainty of the correction is estimated, among others, from
a comparison of \pyth and \herwig. Note that it is important to avoid
double counting of MC uncertainties in the signal MC model and the
JES, as they are both derived in part from \pyth--\herwig comparisons.

Top quark decays are a significant source of $b$~quarks. In general,
jets containing $b$~quarks have a different JES than light parton
jets. The uncertainty of the $b$~JES correction is larger than for
light jets, because there are additional uncertainties due to
fragmentation and color flow, and the small $b$~quark jet sample sizes
make a calibration in data less precise.

In recent years the JES uncertainty has been reduced significantly in
top mass analyses by employing an \emph{in situ} JES calibration
technique~\cite{InSituCDF}. In the $\ttbar\to \ell\nu b\,qqb$ decay
channel, the mass of the top quarks and the mass of the hadronically
decaying $W$ boson are extracted simultaneously. The comparison of the
measured $W$~mass with the world average $W$ mass is used to calibrate
the JES. In this case, the JES uncertainty reduces to a residual
uncertainty that covers the dependence on the jet kinematics.

\subsection{Recent Developments}

The Tevatron collaborations are in the process of revisiting their
procedures for assigning systematic uncertainties. This becomes
important especially for precision measurements of the top quark
mass. Starting from two joint workshops in 2007, CDF and \dzero are
working towards a better understanding of each other's procedures
and a common approach for assigning systematic uncertainties. 

\begin{figure}[t]
  \begin{center}
    \includegraphics[width=0.47\textwidth]{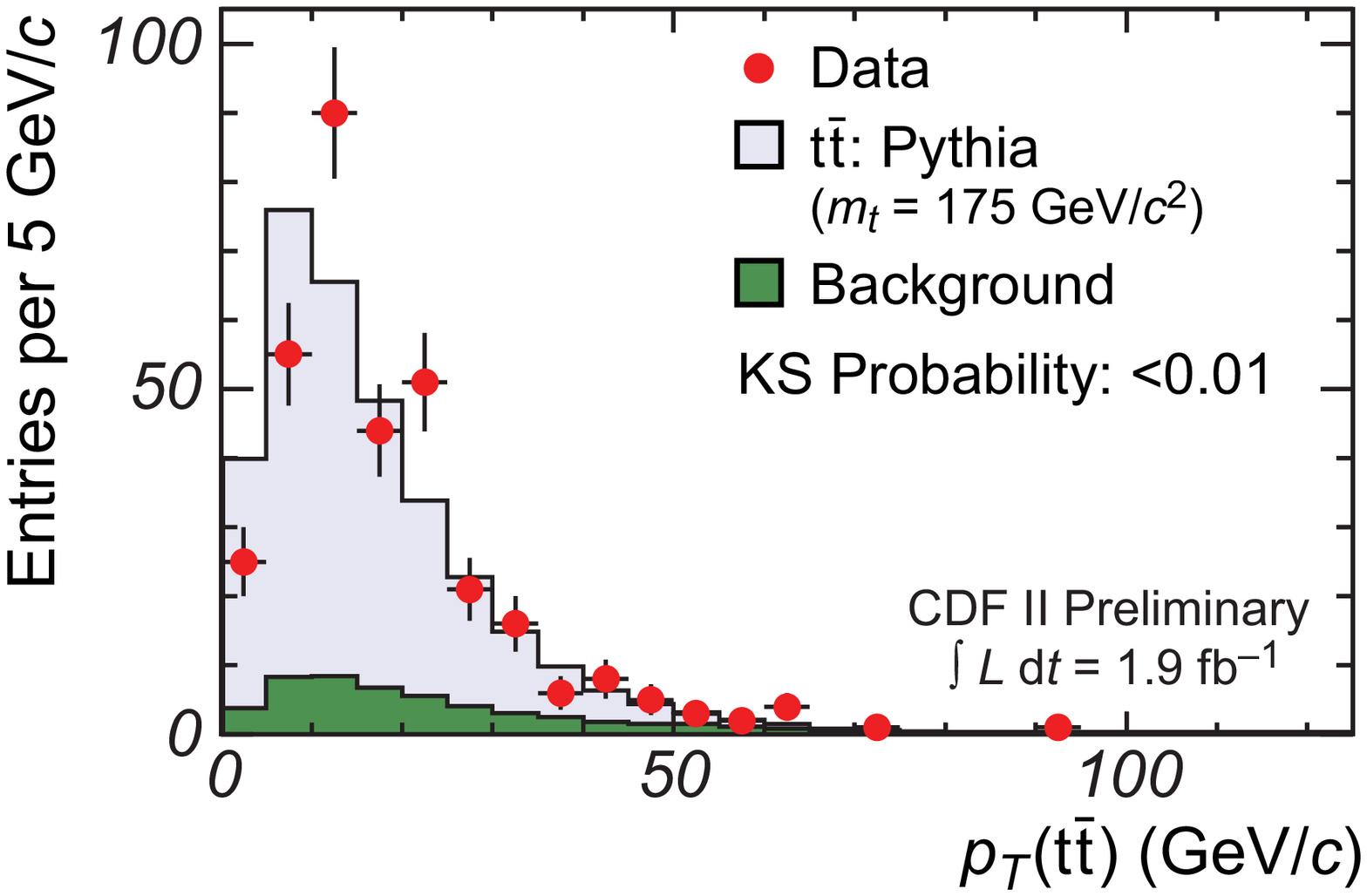}
    \hspace{5mm}
    \includegraphics[width=0.47\textwidth]{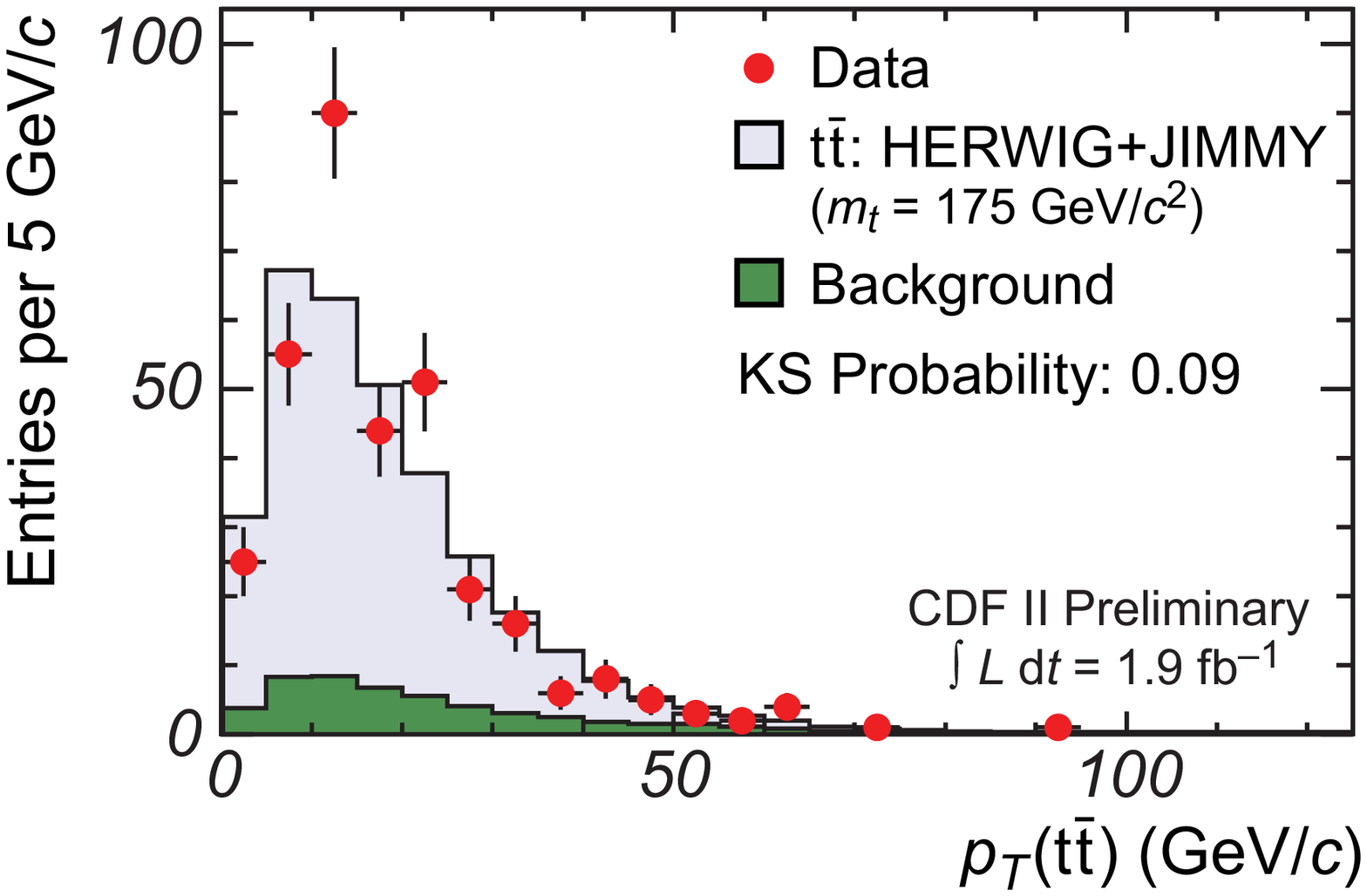}
    \vspace{5mm}

    \includegraphics[width=0.47\textwidth]{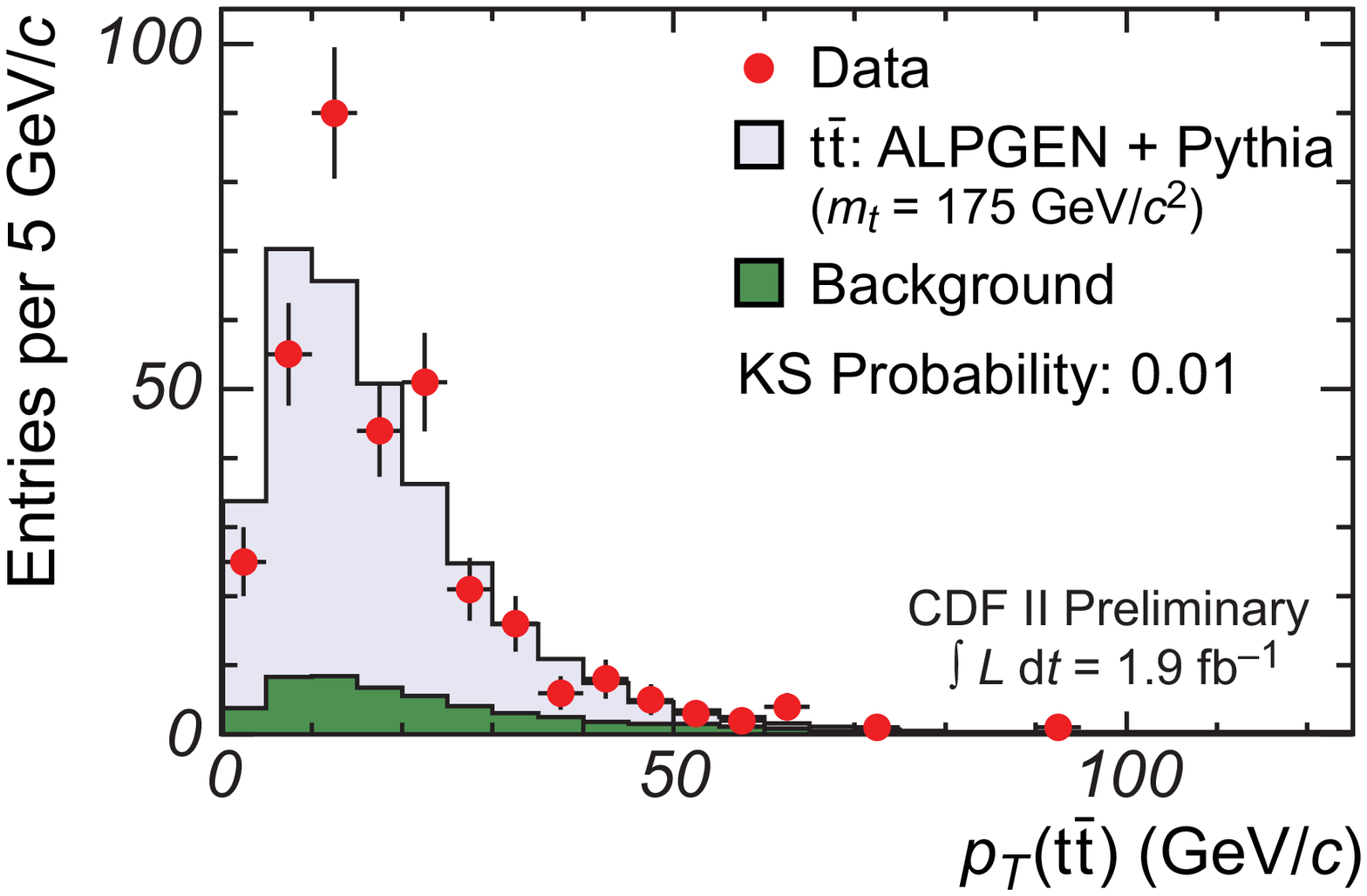}
    \hspace{5mm}
    \includegraphics[width=0.47\textwidth]{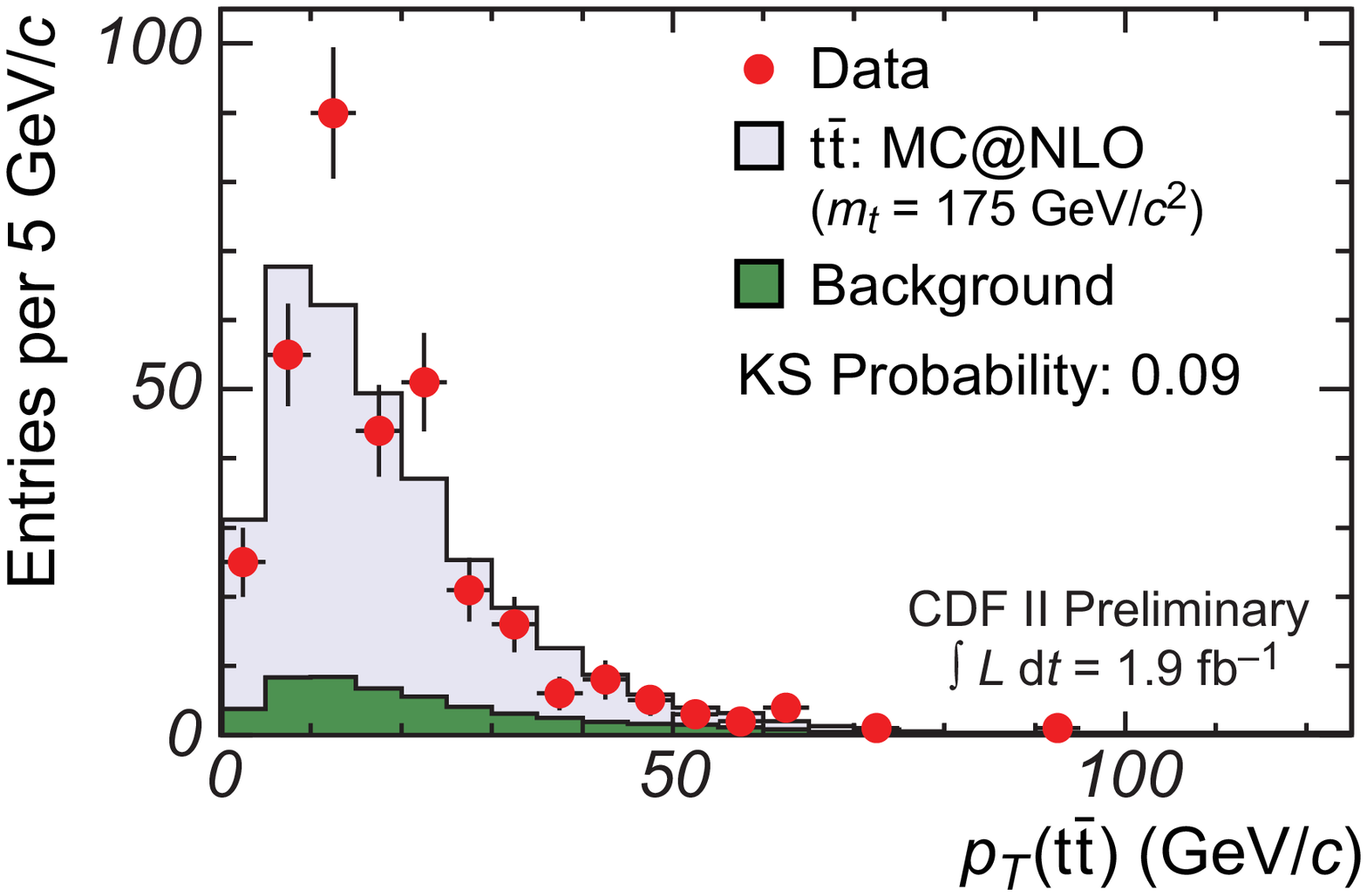}
  \end{center}
  \caption{Comparison of the \ttbar transverse momenta in 1.9\invfb of
    data with different MC generators. Top left: \pyth. Top right:
    \herwig+\,JIMMY. Bottom left: \alp+\,\pyth
    (\ttbar+\,jets). Bottom right: MC@NLO.}
\label{fig:pt_ttbar}
\end{figure}

One example of a long-standing disagreement between Tevatron data and
MC simulations lies in the transverse momentum distribution of \ttbar
pairs. Fig.~\ref{fig:pt_ttbar} shows a comparison of CDF data with
different MC generators. None of the generators shows good agreement
with the data. The \herwig-based generators
(\herwig+\,JIMMY~\cite{JIMMY}, MC@NLO) show better agreement with the
data than the \pyth-based generators (\pyth,
\alp+\,\pyth). Generators that include higher orders
(\alp+\,\pyth, MC@NLO) show the same disagreement as LO generators.
These findings indicate that the source of the disagreement lies in
the PS part of the MC generators. Note that the influence of the
disagreement on the top quark mass uncertainty is small.

%The implications of color reconnection effects on the top mass
%are discussed elsewhere~\cite{SkandsWicke}.

%%%%%%%%%%%%%%%%%%%%%%%%%%%%%%%%%%%%%%%%%%%%%%%%%%%%%%%%%%%%%%%%%%%%%%%%
%
% Special Analyses
%
%%%%%%%%%%%%%%%%%%%%%%%%%%%%%%%%%%%%%%%%%%%%%%%%%%%%%%%%%%%%%%%%%%%%%%%%

\section{Monte Carlo Simulations for Special Analyses}

In many analyses of top quark properties, the standard MC tools for
\ttbar production do not provide all the features required for the
analysis. This is especially true for CDF's main generator \pyth v6.2.
In this section we will discuss examples of Tevatron top properties
analyses with special MC requirements.

\subsection{$W$ Helicity in Top Quark Decays}
\label{section:whel}

Measurements of the helicity of $W$ bosons in top quark decays are
tests of the \VminusA structure of the $tWb$ vertex. The recent
Tevatron analyses~\cite{WhelD0,WhelCDF} measure the $W$ helicity via
the distribution of the angle \costh between the charged lepton (or
down-type quark) and the top boost direction in the $W$ rest frame.

To compare the acceptances for \ttbar processes with different $W$
helicities, the \costh distributions obtained from \alp+\,\pyth
(\dzero) or \pyth (CDF) are reweighted according to helicities other
than the standard model (SM) prediction.  For linearity checks of the $W$
helicity measurement, CDF has also used the \madevent generator, which
changes the underlying left-handed and right-handed couplings rather
than the direct observable \costh.  Systematic uncertainties of the
$W$ helicity measurement are obtained from comparing the SM helicities
predicted by different MC generators, \eg \pyth versus \herwig in the
case of CDF.

\subsection{Flavor Changing Neutral Currents in the Top Sector}

Flavor changing neutral current (FCNC) interactions of top quarks are
strongly suppressed in the SM, so that any FCNC signal would be an
indication for physics beyond the SM. CDF has searched for the FCNC
\tZq~\cite{FCNCCDF} in \ttbar decays, and \dzero has searched for
single top production via the FCNC $q\to tg$~\cite{FCNCD0}.

The collaborations have chosen different approaches to determining the
FCNC signal acceptances. CDF obtains the \tZq decay from \pyth and
reweights the resulting isotropic \costh distribution (defined
analogously to sect.~\ref{section:whel}) according to the
expectations. This approach stays close to experimental observables,
which can also be seen by the result, which is a limit on the
branching fraction for the \tZq decay. The \dzero analysis utilizes
the CompHEP MC generator~\cite{CompHEP} to modify the top quark
couplings. The \dzero approach is closer to theoretical calculations,
and consequently, the result is quoted as a limit on the $qtg$ coupling.

\subsection{Fraction of \ttbar Events Produced in Gluon-Gluon Fusion}

NLO QCD calculations predict that at the Tevatron, 85\% of all \ttbar
pairs are produced via \qqbar annihilation and 15\% are produced via
$gg$ fusion. CDF has recently published first experimental tests of
this prediction~\cite{ggFraction}. One of the CDF analyses measures
the fraction of \ttbar produced via $gg$ fusion by training an
artificial neural network on the kinematics of \ttbar production and
decay. While the main sensitivity of this method comes from the
production angles and velocities of the top quarks, the dependence on
\ttbar spin correlations, which is encoded in the decay kinematics, also
contributes to the sensitivity.

\begin{figure}[t]
  \begin{center}
    \includegraphics[width=0.47\textwidth]{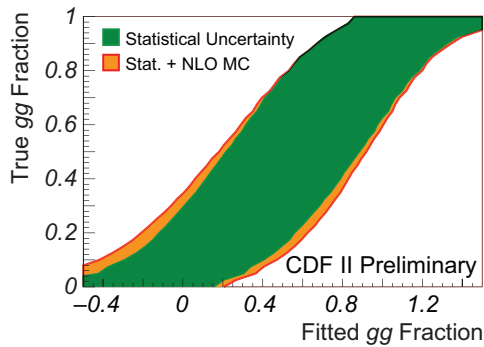}
  \end{center}
  \caption{Feldman-Cousins (FC) bands at 68\% confidence level for the
    measurement of the fraction of \ttbar production via $gg$ fusion.
    The FC band for statistical uncertainties plus uncertainties due to
    the comparison of \herwig with MC@NLO is compared with the FC band 
    for statistical uncertainties only. At the fitted $gg$ fraction
    of $-7.5\%$, the limit on the true $gg$ fraction is deteriorated by
    approximately 20\%.}
\label{fig:ggfraction}
\end{figure}

The standard CDF MC tool \pyth is missing two important ingredients
for this analysis. \pyth as a LO MC generator only generates 5\%
$gg\to\ttbar$, and it does not include \ttbar spin correlations. Therefore
the analysis makes use of the \herwig generator, also using a
CDF-internal extension of \herwig named GGWIG that allows the user to
adjust the relative fraction of $gg\to\ttbar$. The systematic
uncertainty of the measurement due to the difference between \herwig
and \pyth is small; however, the systematic difference between \herwig
and the NLO generator MC@NLO results in a 20\% loss in sensitivity,
see fig.~\ref{fig:ggfraction}. This is one of the largest systematic
uncertainties and indicates the importance of NLO effects for the
analysis.

\subsection{Charge (Forward Backward) Asymmetry}

Both Tevatron collaboration have recently published measurements of
the charge asymmetry in \ttbar production (which translates into a
forward-backward asymmetry if one assumes CPT
invariance)~\cite{FBD0,FBCDF}. The asymmetry vanishes at LO, and NLO
calculations predict an asymmetry of 4--5\%~\cite{Kuehn}, which is
confirmed by the MC@NLO generator, which shows a 3.8\% asymmetry.

\begin{figure}[t]
  \begin{center}
    \includegraphics[width=0.51\textwidth]{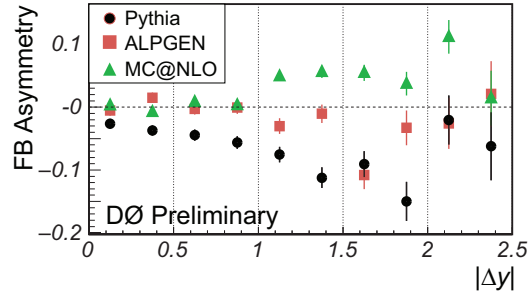}
  \end{center}
  \caption{Charge (forward-backward) asymmetry as a function of the
    rapidity difference between the $t$ and the $\tbar$ quark for the
    \pyth, \alp, and MC@NLO generators.}
\label{fig:fb}
\end{figure}

The net asymmetry obtained from LO MC generators is zero; however,
there may be asymmetries generated from color flow for specific
kinematic configuration.  This is illustrated in fig.~\ref{fig:fb},
which shows a comparison of the asymmetry as predicted in \pyth, \alp,
and MC@NLO as a function of the rapidity difference between the $t$
and the $\tbar$ quark.  The asymmetries are also subject to NLO
corrections. For example, for $\ttbar\,+$\,jet production, a LO
asymmetry of $-8\%$ is reduced to $-1.5\%$~\cite{Dittmaier}. In
summary, there is currently no single MC generator that covers all
features of the charge (FB) asymmetry.

%%%%%%%%%%%%%%%%%%%%%%%%%%%%%%%%%%%%%%%%%%%%%%%%%%%%%%%%%%%%%%%%%%%%%%%%
%
% Summary
%
%%%%%%%%%%%%%%%%%%%%%%%%%%%%%%%%%%%%%%%%%%%%%%%%%%%%%%%%%%%%%%%%%%%%%%%%

\section{Summary and Outlook}

The CDF and \dzero collaborations have entered an era of precision top
quark physics. While more and more sophisticated analysis techniques
are being employed, both collaborations favor a ``conservative'' and
``pragmatic'' approach towards MC simulations:
\begin{itemize}
\item Prefer well validated MC tools (mainly based on \pyth)
  over ``cutting edge'' MC technology.
\item Prefer data-driven methods whenever available.
\item Adapt and re-use well understood MC samples, \eg by applying
  reweighting techniques.
\end{itemize}

In addition there are often technical reasons for the above
approach. Integrating new MC generators into the existing CDF and
\dzero frameworks is time intensive and error prone, despite close
collaboration with the authors of the MC~codes.  With the current
computing and person power available at the Tevatron, the cycle of
software integration, event generation, validation, and tuning of a
new MC generator can easily take many months.

The Tevatron collaborations are actively working on common approaches
to top MC generators. The final goal is a common treatment of
MC-related systematic uncertainties, primarily for the top quark mass
measurements, but also for other top analyses.

%%%%%%%%%%%%%%%%%%%%%%%%%%%%%%%%%%%%%%%%%%%%%%%%%%%%%%%%%%%%%%%%%%%%%%%%
%
% Acknowledgments
%
%%%%%%%%%%%%%%%%%%%%%%%%%%%%%%%%%%%%%%%%%%%%%%%%%%%%%%%%%%%%%%%%%%%%%%%%

\acknowledgments It is a pleasure to thank the organizers of TOP2008
for an extremely productive workshop at the perfect venue. This work was
supported by the Helmholtz Association under contract number
VH-NG-400.

%%%%%%%%%%%%%%%%%%%%%%%%%%%%%%%%%%%%%%%%%%%%%%%%%%%%%%%%%%%%%%%%%%%%%%%%
%
% Bibliography
%
%%%%%%%%%%%%%%%%%%%%%%%%%%%%%%%%%%%%%%%%%%%%%%%%%%%%%%%%%%%%%%%%%%%%%%%%


\begin{thebibliography}{99}

\bibitem{TopMass2008}
  \BY{Tevatron Electroweak Working Group}
  arXiv:0803.1683 [hep-ex].

\bibitem{SingleTopD0}
  \BY{Abazov V. M. \etal{} (\dzero Collaboration)}
  \IN{Phys. Rev. Lett.}{98}{2007}{181802}.

\bibitem{SingleTopCDF}
  \BY{CDF Collaboration}
  CDF Public Notes 9217, 9221, 9223, 9252.

\bibitem{Pythia}
  \BY{Sj{\"o}strand T. \etal}
  \IN{Comput. Phys. Commun.}{135}{2001}{238}.

\bibitem{ALPGEN}
  \BY{Mangano M.~L. \etal}
  \IN{JHEP}{07}{2003}{001}.

\bibitem{CTEQ5}
  \BY{Lai H. L. \etal{} (CTEQ Collaboration)}
  \IN{Eur. Phys. J. C}{12}{2000}{375}.

\bibitem{CTEQ6}
  \BY{Pumplin J. \etal{} (CTEQ Collaboration)}
  \IN{JHEP}{07}{2002}{012}.

\bibitem{ZTOP}
  \BY{Sullivan Z.}
  \IN{Phys. Rev. D}{70}{2004}{114012}.

\bibitem{Boos}
  \BY{Boos E.~E. \etal}
  \IN{Phys. Atom. Nucl.}{69}{2006}{1317}.

\bibitem{Lueck}
  \BY{L\"uck J.}
  FERMILAB-MASTERS-2006-01.

\bibitem{TeV4LHC}
  \BY{Gerber C. E. \etal{} (TeV4LHC-Top and Electroweak Working
    Group)}
   arXiv:0705.3251 [hep-ph].

\bibitem{MadEvent}
  \BY{Maltoni F. \atque Stelzer T.}
  \IN{JHEP}{02}{2003}{027}.

\bibitem{MCATNLO}
  \BY{Frixione S., Nason P. \atque Webber B. R.}
  \IN{JHEP}{03}{2003}{007}.

\bibitem{Herwig}
  \BY{Corcella G. \etal}
  \IN{JHEP}{01}{2001}{010}.

\bibitem{MRST}
  \BY{Martin A. D., Roberts R. G., Stirling W. J. \atque Thorne R. S.}
  \IN{Eur. Phys. J. C}{4}{1998}{463}.

\bibitem{JESCDF}
  \BY{Bhatti A. \etal}
  \IN{Nucl. Instrum. Meth. A}{566}{2006}{375}.

\bibitem{InSituCDF}
  \BY{Abulencia A. \etal{} (CDF Collaboration)}
  \IN{Phys. Rev. D}{73}{2006}{032003}.

\bibitem{JIMMY}
  \BY{Butterworth J. M., Forshaw J. R. \atque Seymour M. H.}
  \IN{Z. Phys. C}{72}{1996}{637}.

\bibitem{WhelD0}
  \BY{Abazov V. M. \etal{} (\dzero Collaboration)}
  \IN{Phys. Rev. Lett.}{100}{2008}{062004}.

\bibitem{WhelCDF}
  \BY{CDF Collaboration}
  CDF Public Notes 9114, 9144, 9215.

\bibitem{FCNCCDF}
  \BY{Aaltonen T. \etal{} (CDF Collaboration)}
  arXiv:0805.2109 [hep-ex].

\bibitem{FCNCD0}
  \BY{Abazov V. M. \etal{} (\dzero Collaboration)}
  \IN{Phys. Rev. Lett.}{99}{2007}{191802}.

\bibitem{CompHEP}
  \BY{Pukhov A. \etal}
  hep-ph/9908288.

\bibitem{ggFraction}
  \BY{CDF Collaboration}
  CDF Public Note 8811.

\bibitem{FBD0}
  \BY{Abazov V. M. \etal{} (\dzero Collaboration)}
  \IN{Phys. Rev. Lett.}{100}{2008}{142002}.

\bibitem{FBCDF}
 \BY{Aaltonen T. \etal{} (CDF Collaboration)}
 arXiv:0806.2472 [hep-ex].

\bibitem{Kuehn}
  \BY{K\"uhn J. H. \atque Rodrigo G.}
  \IN{Phys. Rev. Lett.}{81}{1998}{49}.

\bibitem{Dittmaier}
  \BY{Dittmaier S., Uwer P. \atque Weinzierl S.}
  \IN{Phys. Rev. Lett.}{98}{2007}{262002}.

\end{thebibliography}
\end{document}